# Thermal neutron radiography of a passive proton exchange membrane fuel cell for portable hydrogen energy systems

Antonio M. Chaparro[(1),*], P. Ferreira-Aparicio[(1)], M. Antonia Folgado[(1)],

Rico Hübscher[(2)], Carsten Lange[(2)], Norbert Weber[(3)]

(1) Dep. of Energy. CIEMAT. Avda. Complutense, 40. 28040 Madrid. Spain.

(2) TU Dresden. Faculty of Mechanical Science and Engineering. Institute of Power Engineering. 01062 Dresden. Germany.

(3) Institut für Fluiddynamik. Helmholtz-Zentrum Dresden-Rossendorf e.V. Bautzner Landstr. 400, 01328 Dresden. Germany.

*Corresponding author:

Antonio M. Chaparro

Energy Department, CIEMAT, Avda. Complutense 40. 28040 Madrid, Spain

antonio.mchaparro@ciemat.es

Telephone: +34 913460897

Fax: +34 913466269

**Abstract**

A proton exchange membrane fuel cell (PEMFC) for low power and portable applications is studied with thermal neutron radiography. The PEMFC operates under full passive conditions, with an air-breathing cathode and a dead-end anode supplied with static ambient air and dry hydrogen, respectively. A columnar cathodic plate favors the mobility of water drops over the cathode surface and their elimination. Thermal neutron images show liquid water build up during operation with the cell in vertical and horizontal positions, i.e. with its main plane aligned parallel and perpendicular to the gravity field, respectively. Polarization curves and impedance spectroscopy show cell orientation dependent response that can be related with the water accumulation profiles. In vertical position, lower water contents in the cathode electrode is favored by the elimination of water drops rolling over the cathode surface and the onset of natural convection. As a consequence, oxygen transport is improved in the vertical cell, that can be operated full passive for hours under ambient conditions, providing steady peak power densities above 100 mW $cm^{-2}$. Gravity and natural convection are less effective in horizontal position, leading to a 17 % decrease in peak power density due to oxygen transport losses. The horizontal position is especially adverse if the upper electrode is the cathode, because of anode flooding causing cell failure after production of a small amount of water (5 mg $cm^{-2}$). The combined information from thermal neutron radiography and cell response characteristics explains the important influence of cell orientation on the performance of a fully passive air-breathing PEMFC for portable applications.

# 1. Introduction

Over the last two decades, the development of portable hydrogen fuel cell systems has attained sufficient maturity for low power and portable applications which benefit from their high power density and safety [1,2,3,4,5,6,7]. Particular operation conditions in portable applications dictate changes in the design of the cells due to autonomy, compactness, and light-weight power production requirements, as well as their proximity to the user should not compromise personal comfort and safety, by emission of heat, liquid water, moving parts, or noises [8].

A *passive air-breathing cathode* is most suitable for a portable proton-exchange membrane fuel cell (PEMFC) working with ambient air and passive rejection of water [9,10,11,12]. No applied convective force is required for oxygen and water transports, but only natural forces originating from concentration and heat gradients, water evaporation-condensation, and gravity drag of water drops over the cathode surface. Portability of a PEMFC is improved if using a *passive anode* in front of the air-breathing cathode. The passive anode operates in dead-end mode in a static atmosphere of dry hydrogen, without periodical purging, which improves the efficiency and autonomy of the power system. For continuous passive operation, the flooding and drying-out of cathode and anode must be prevented, that depend on a delicate equilibrium between different operation parameters [13,14]. Incorporating passive water rejection means, like superhydrophobic layers that repeal the liquid water from porous electrodes [15], or a hydrophilic membrane to allow water permeation from the anode [16] help the passive PEMFC operation, where water management is more critical than for a conventional, flow-field based, PEMFC.

A technique that provides most useful information of water dynamics in a PEMFC is *neutron radiography*. Neutrons have a high sensitivity to liquid water inside porous structures and solid volumes. High lateral and temporal resolutions can be attained, with pixel sizes as low as 30 µm and image acquisition times from 5 s. Studies in-operando have been carried out with conventional [17,18,19,20,21,22], and passive air-breathing PEMFC [10,23]. In conventional cells, water accumulation is observed in the channels of the flow-field plates for different designs and operation conditions. Siegel et al. [17] showed water accumulation and voltage drop in the anode and cathode of a cell fed with dry hydrogen and periodical purging. At high current densities, neutron images show intense water accumulation under the ribs of flow-field plates leading to transport losses parallel and perpendicular to the cell plane dependent on cell compression [18,21]. The influence of the gas flow configuration in a water cooled commercial stack was studied by Iranzo et al. by combining neutron radiography with localized temperature, current density, and modelling [21]. Neutron imaging has also been used for the observation of water inside the gas diffusion layer by dark field technique with a grating interferometer [24]. A few studies with neutron imaging have been carried out on passive air-breathing cells. Weiland et al. studied cells changing the configuration of the cathodic plate, particularly the number and size of openings [10]. They found accumulation of the water produced by the cell only in the cathode, not in the anode, and attribute voltage losses to the obstruction of the openings. Coz et al. [23] studied water management in a planar air-breathing stack fabricated on printed circuit board, and delivering up to 150 mW cm$^{-2}$; their cell design showed heterogeneous water accumulation in the anode and cathode leading to 50% loss in performance after ten hours of operation. In addition to neutron radiography, other techniques used for in-operando liquid water studies of PEMFCs are the X-ray radiography [25,26,27,28,29,30], droplet microvisualization inside optically transparent cells [31], and nuclear magnetic resonance [32], some of them requiring changes in cell configuration that may alter operation characteristics [33].

In this work, thermal neutron radiography is used for the in-operando study of a fully passive PEMFC for portable applications. Thermal neutrons, produced by nuclear fission and a moderator material to slow down, are used for the first time with this research objective. The PEMFC design is based on the concept depicted in **Fig. 1**, characterized by passive anode and cathode able to operate under static hydrogen and air atmospheres [34,35]. Neutron imaging experiments have been performed keeping the original design almost unmodified, except for the use of aluminum anodic plates, which are almost transparent to thermal neutrons, instead of plastic plates. Cells are studied in vertical and horizontal positions to observe how natural forces determine passive water transport and performance. Results are complemented with polarization curves and impedance spectroscopy.

## 2. Experimental

*2.1 Passive PEMFC description*

Characteristics of the passive PEMFC design are shown in **Fig. 1a**, and the scheme of operation in **Fig. 1b**. The cell has a circular design with an active area of 7.1 cm$^2$, and optimized anode and cathode architectures for passive operation. The anodic plate is a perforated disk made of aluminum alloy (SIMAGALTOK 82 EN AW 6082, 5mm thickness, 5cm diameter) that allocates a gas diffusion electrode (GDE) (ELAT GDE LT250EWALTSI, BASF, 0.25 mg$_{Pt}$ cm$^{-2}$) and an Au covered Ni grid (Dexmet) current collector. For gas tightness, the collector grid is framed with 3D printed PLA and sealed with silicone gaskets. A hydrophilic membrane (Nafion 117, Ion Power Inc.) is placed at the back of the perforated anodic plate to help for the passive liquid water removal from the anode (Fig. 2b). The cathode is open air, with a columnar plate (adapted from an aluminum pin heatsink, Fischer Elektronik), a grid current collector (Au covered Ni grid from Dexmet), and a gas diffusion electrode (W1S1009, FuelCellsEtc., 0.30 mg$_{Pt}$ cm$^{-2}$). The electrolyte is a Nafion 212NR membrane (Ion Power Inc.). All cell elements, except the cathodic plate, are fixed with eight stainless steel bolts (M3) and nuts at 1 N m torque; the cathodic plate is tightened independently to the cell body with another set of nuts at 0.5 N m torque.

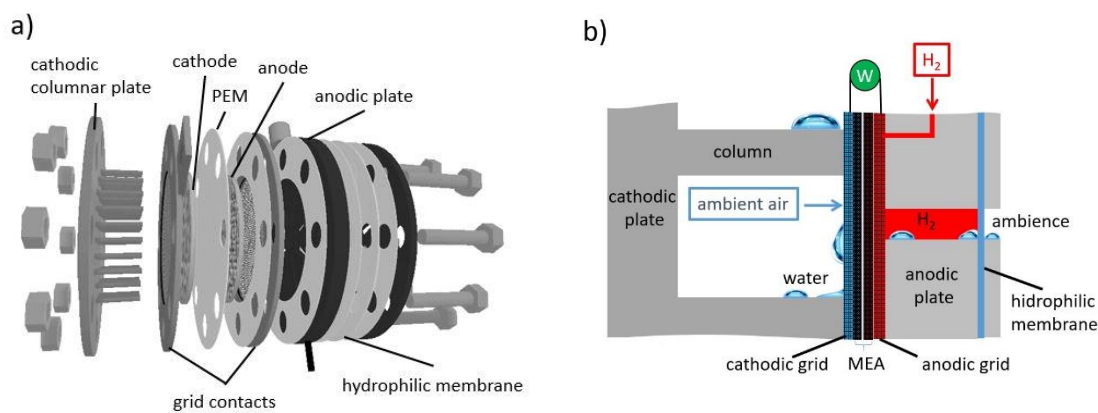

**Fig. 1.** a) Exploded view of the passive PEMFC components; b) scheme of the passive operation.

*2.2 Electrochemical characterization*

The cell response was studied with polarization curves and impedance spectroscopy, in three cell orientations: vertical, horizontal with cathode up, and horizontal with cathode down. Cells were supplied with 0.5 $bar_g$ static hydrogen (Air Liquide, 99.999 %) in the anode, and static ambient air, at 22 °C and 30 % RH inside a climatic chamber (Dycometal), in the cathode. Polarization curves were obtained potentiostatically (Autolab 30 N with 10 A current booster), by stepping cell voltage from open circuit to 0.3 V, at 10 mV $s^{-1}$ and 2.4 mV steps. Before the acquisition of the curves, cells were brought to and kept in a steady-state by polarizing at 0.5 V during 30 min until constant current and temperature were attained. During the measurement of the polarization curve, the ventilator of the climatic chamber was switched-off to allow for a fully static ambient air. The temperature of the cells was monitored during measurements by means of a thermocouple put in contact with the cathodic plate.

Impedance spectroscopy measurements were carried out potentiostatically (sine wave, $V_{ac}$ = 10 $mV_{RMS}$) from 20 kHz to 0.1 Hz (50 points), with the cells placed inside the climatic chamber and ventilator switched-off during measurements. Previous to impedance acquisition, the cell is polarised at the predetermined potential until its temperature and current response are stabilized. The impedance spectra were submitted to Kronig-Kramer tests and analyzed by fitting to an equivalent circuit using Autolab software (Nova). Kronig-Kramer tests, to check stability and linearity conditions, were satisfactorily passed for all spectra.

*2.3 Neutron radiography*

*2.3.1 Thermal neutron source and imaging set-up.* Neutron imaging of an in-operando fuel cell with thermal neutrons was performed for the first time at the training and research reactor AKR-2 of the Technische Universität Dresden [36]. The AKR-2 is a very low flux, thermal, homogeneous, solid moderated zero-power reactor with a maximum continuous power of 2 W. Nuclear fission is mainly caused by thermal neutrons, i.e. neutrons with the same kinetic energy as the particles of ambient materials. The nuclear fuel and moderator material (solid polyethylene) is distributed homogeneously in fuel plates. Radial channel 7 was used for best combination of exposure time and effective beam divergence, with an absolute thermal neutron flux of $\Phi$ = 2300 n cm$^{-2}$ s$^{-1}$ (±5%) [36]. The raw data of the neutron radiographs were taken with an upgraded system containing an Andor iKon-M 934 (1024x1024 pixels) camera cooled down to -70 °C combined with a light tight DNIDS box provided by Paul-Scherrer-Institute, a Nikkor AF-S 50 mm 1:1,4G lens and a 200 µm thick $^6$LiF/ZnS(Ag) scintillator from Tritec. The effective pixel size has been estimated to be about 0.133 mm, which results in a spatial resolution of 0.266 mm, that may have additional influences, like scattering terms. See Supplementary Information, S1.1, for an experimental test of the spatial resolution. More details of the neutron radiography installation and imaging set-up will be presented in a forthcoming publication.

*2.3.2 Fuel cell installation and operation.* The passive PEMFC is fixed to an aluminum holder in front of the radial channel, with the possibility for vertical and horizontal cell positions, i.e. main plane parallel and perpendicular to the gravity field, respectively. The vertical cell was studied under frontal and lateral irradiation. During acquisition of the neutron radiography, the cell current is controlled (Keithley 2460 Sourcemeter), while supplied with dry hydrogen at 0.5 bar$_g$ static pressure from a metal-hydride canister (Horizon Hydrostick Pro, 1 g H$_2$), and the static ambient air (22 °C, 30 % RH). Cell voltage and temperature are continuously recorded using a Keithley 2701 Multimeter with a 7700 multiplexer card. **Figs. S1a** and **b** of the Supplementary Information shows photographs of the set-up. Four cell configurations have been radiographed: vertical cell with frontal (1), and lateral (2) incidence, and horizontal cell with anode on top (3), and cathode on top (4). Fig. S1b shows the cell in configuration (1).

*2.3.3. Image acquisition and processing.* All neutron images were corrected for the dark-field and the flat-field. Dark-field correction consists of eliminating the pixel bias and the time-dependent dark current noise. With this aim, a series of images is taken with the camera shutter permanently closed. The recorded images are averaged resulting in the dark-field master image (*DF*). Correction for the neutron beam profile is handled via flat-fielding, by taking a series of images without any object in-between the neutron source and the detector and averaging them (*FF*). The final image (*IM*) is obtained from the corrected neutron images (*NI*) according to:

$$IM = \frac{NI-DF}{FF-DF} \, m_{FF} \qquad \text{Eq. 1}$$

where $m_{\text{FF}}$ is the calculated median (middle value separating the higher half from the lower half pixel sets) of the *FF* image. Averaging of a series of sub-exposure images, including *DF* and *FF*, was always performed by the median pixel of the series and without a 2D-Median Filter.

Neutron images were taken in-operando at different current demands. In each experiment, before starting images acquisition, the cell was dried by thorough spraying of compressed air over the cathode surface and through the anode chamber. The dried cell was then put in the holder and the dry-cell reference (*IM$_{dry}$*) image obtained from five sub-exposure images, and after correction (Eq. 1). Then, the cell was started-up by increasing progressively the current demand in steps and simultaneous neutron images acquisition. Acquisition time for neutron image was 202 s (180 s sub-exposure and about 22 s readout time).

Images of the generated liquid water ($IM_W$) were obtained from the corrected fuel cell images ($IM$), by dividing by the reference, 'dry-cell', image ($IM_{dry}$), and taking the natural logarithm:

$$IM_W = \log\left(\frac{IM}{IM_{dry}}\right) \qquad \text{Eq. 2}$$

To improve the signal-to-noise ratio and match the exposure time with $IM_{dry}$, moving averages of five consecutive images were computed. All digital image processing was carried out with the publicly available program ImageJ. Additional information of the neutron imaging characteristics is given in Supplementary Information (S1), including the analysis of gray scale as a function of water produced in the cell, and the spatial resolution of the images.

### 3. Results

*3.1 Polarization curves and impedance spectroscopy*

*3.1.1 Polarization curves.* Polarization curves of the passive PEMFC are shown in **Fig. 2** for three cell orientations: vertical, horizontal cathode-down, and horizontal cathode-up. The curves were taken after three cycles from open circuit to 0.3 V, at 22 °C and 30 % RH under ambient static air, while feeding the dead-end anode with static dry $H_2$ (pressure 0.5 bar$_g$). The temperature of the cells during polarization curves acquisition was within 25 - 27 °C (see Supplementary Information, S2, Fig. S4), which is a result of its heat generation and dissipation to the surroundings. The curves show the best performance for the vertical cell with a maximum power density of 90 mW cm$^{-2}$, compared with 75 mW cm$^{-2}$ for the horizontal cell. Little difference is observed between the cathode-up or -down in the horizontal cell. However, cathode-up leads to cell failure in constant current experiments (see next section).

Polarization curves have been analyzed with a model for passive air-breathing PEMFC which accounts for changes in oxygen diffusivity inside the catalyst layer and the gas diffusion layer of the PEMFC cathode [37]. Details of the model, assumptions, and calculations are given in Appendix 1. Results of the fitting to the theory curve show that cell orientation is accompanied by a change in oxygen diffusivity of the cathodic catalyst layer, from $8.0 \cdot 10^{-4}$ cm$^2$ s$^{-1}$, to $6.7 \cdot 10^{-4}$ cm$^2$ s$^{-1}$, and $6.5 \cdot 10^{-4}$ cm$^2$ s$^{-1}$, for the vertical, horizontal with cathode down, and horizontal with cathode up cell, respectively. In the gas diffusion layers, the calculated oxygen diffusivities are $145 \cdot 10^{-3}$ cm$^2$ s$^{-1}$, $109 \cdot 10^{-3}$ cm$^2$ s$^{-1}$, and $111 \cdot 10^{-3}$ cm$^2$ s$^{-1}$ for vertical, horizontal cell with cathode down, and cathode up, respectively. The diffusivity values are gathered in **Table I**.

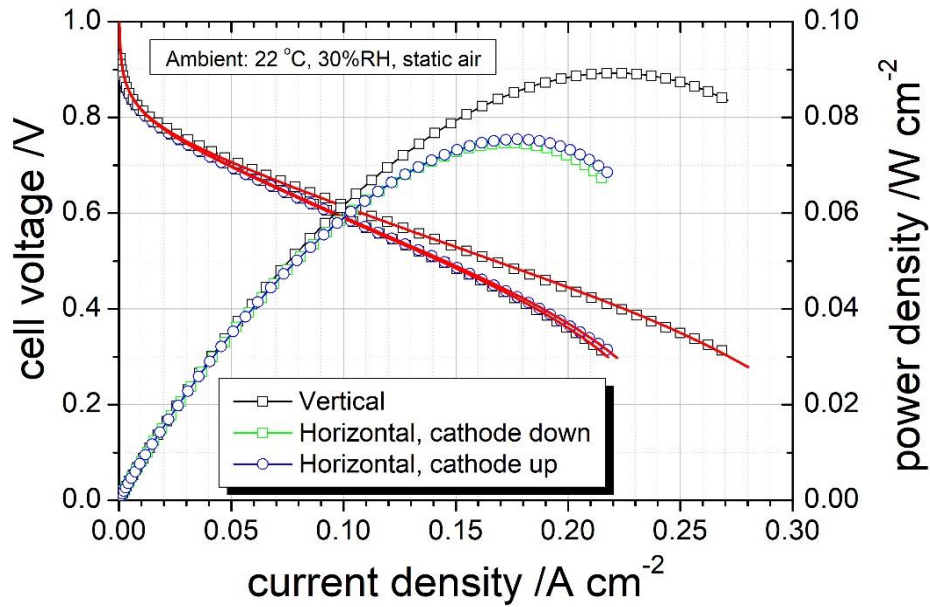

**Fig. 2.** Polarization curves and power density of the passive PEMFC in three different orientations. Cell supplied with dry H$_2$ in dead-end mode, and static ambient air. Pt loadings of anode/cathode: 0.25/0.30 mg cm$^{-2}$. Red line in polarization curves correspond to the best fit to the theory curve (Appendix 1).

Oxygen diffusivity changes of the porous layers can be related with changes in their water saturation (*s*), by using the expression [38]:

$$D_{eff} = D_{bulk} f(\varepsilon) g(s) \qquad \text{Eq. 3}$$

where $D_{eff}$ and $D_{bulk}$ are the effective diffusivity and the open space diffusivity, respectively, $\varepsilon$ the porosity, $f(\varepsilon)$ is the ratio of the dry diffusivity of the film and $D_{bulk}$, and $g(s)$ is the ratio of $D_{eff}$ and dry diffusivities of the film (dry diffusivity of the porous layer cancels out therefore in Eq. 3). Different expressions of $f(\varepsilon)$ and $g(s)$ have been proposed for the catalyst layer and the gas diffusion layer of a PEMFC electrode [38,39,40,41,42,43,44]. Taking for the gas diffusion layer the expression by García-Salaberri et al. [38], and for the catalyst layer that by Zhen and Kim [44], the resulting water saturations for the layers of the cathode are given in Table I. Calculations assume laterally homogeneous layers, with the microporous carbon layer, present between the catalyst and the gas diffusion layers, included in the second one. Water saturation values in Table I show 35 % larger water content in the gas diffusion layer of the horizontal cell, while the catalyst layer saturation does not change much with cell orientation. It is also remarkable that the two horizontal positions, cathode-up and -down, present similar water saturation conditions in the two layers.

| **Cell position** | Vertical | Horizontal cathode-down | Horizontal cathode-up |
|---|---|---|---|
| $D_{eff,GDL}$ /$10^{-4}$ cm$^2$ s$^{-1}$ | 145 | 109 | 111 |
| $s_{avg,GDL}$ (1) | 0.38 | 0.59 | 0.58 |
| $D_{eff,CL}$ /$10^{-4}$ cm$^2$ s$^{-1}$ | 8.0 | 6.7 | 6.5 |
| $s_{avg,CL}$ (2) | 0.71 | 0.73 | 0.73 |

(1) $f(\varepsilon)=\varepsilon^{3.5}$; $g(s) = (1-s)^{2.15}$ (ref. 38)

(2) $f(\varepsilon)=1.07\varepsilon^{1.75}$; $g(s) = (1-s)^3$ (ref. 44)

Table I. Diffusivities of the cathodic gas diffusion layer ($D_{eff,GDL}$) and catalyst layer ($D_{eff,CL}$), obtained from the analysis of polarization curves (Appendix 1), and water saturations ($s_{avg,GDL}$, $s_{avg,CL}$) calculated from Eq. 3, using expressions in foot notes (1) and (2) and $D_{bulk}=D_{O2-N2} = 0.20$ cm$^2$ s$^{-1}$ (1 atm, 20 ºC) [45]; $\varepsilon_{GDL} = 0.75$; $\varepsilon_{CL} = 0.35$.

The analysis of polarization curves shows that cell orientation changes water saturation conditions in the cathode, mostly in its gas diffusion layer, which reduces oxygen diffusivity and leads to a 17 % decrease in peak power density for the horizontal with respect to the vertical cell. Impedance spectroscopy provides more information on transport limitations arising from the change in cell orientation.

*3.1.2 Impedance spectroscopy.* Results of impedance analysis are shown in Fig. S5 of Supplementary Information, S3, in the form of plots of the imaginary part as a function of the real part of the impedance, or Nyquist plots (-Z'' vs. Z'), at two cell voltages and same cell orientations as in Fig. 2. A flattened single semicircle is observed in all cases which is due to the cathodic impedance of the cell [46]. Analysis of the spectra was carried out by fitting to an equivalent circuit (inset Fig. S5a), with a serial resistance ($R_s$) that accounts for high frequency ohmic losses, i.e. the ionic resistance of membrane and electrodes and electronic resistances, and a parallel combination of a resistor ($R_1$) and constant phase element ($Y_1$, $n$) that accounts for the low frequency losses. Resulting values of the fitting are given in **Table II**, together with the steady-state cell temperature during the measurements. They show a 40 % increase in $R_1$, the low frequency resistor by changing from vertical to horizontal cell position, at 0.3 V, and a minor increase at 0.7 V (6 %). This component of the equivalent circuit reflects the limitations for current generation by slow processes, which for a PEMFC under normal operation conditions correspond either to oxygen reduction in the cathodic catalyst layer, and/or to oxygen transport impedance influenced by the presence of liquid water in the gas diffusion layer or the flow-field channels [46]. Since oxygen reduction kinetics accelerates at low cell voltage, due to favored charge transfer with increasing overvoltage, while the reverse trend is observed in Table II, it can be inferred that $R_1$ is mostly determined by increasing mass transport losses in the passive PEMFC cathode when changing from vertical to horizontal position.

Additional information from the impedance spectroscopy can be obtained from other parameters in Table II. The constant phase element, which reflects the charge storage capacitance of carbon and platinum in the cathodic catalyst layer, shows little variation with cell orientation. In fact, the associated effective capacitance, $C_1^*$, calculated from $Y_1$, $n$ and $R_1$ (Table II) [47,48], which is proportional to the electrochemically accessible electrode area, remains almost unaltered by changing cell orientation. Finally, a slight increase by 10 % in the serial resistance ($R_s$) for the horizontal cells must be attributed to 3-4 °C lower self-heating steady-state temperature attained by the cell.

| Cell Orientation | Vertical | | Horizontal cathode down | | Horizontal cathode up | |
|---|---|---|---|---|---|---|
| Cell voltage /V | 0.7 | 0.3 | 0.7 | 0.3 | 0.7 | 0.3 |
| Cell temp. / °C [1] | 29 | 34 | 26 | 30 | 26 | 30 |
| $R_s$ /mOhm cm$^2$ | 550 (±4) | 503 (±5) | 595 (±4) | 551 (±5) | 589 (±4) | 539 (±6) |
| $R_1$ /mOhm cm$^2$ | 1753 (±14) | 2121 (±35) | 1866 (±14) | 2962 (±35) | 1894 (±14) | 3032 (±42) |
| $Y_1$ /mF s$^{n-1}$ cm$^{-2}$ | 25 (±1) | 45 (±2) | 22 (±1) | 32 (±1) | 23 (±1) | 32 (±1) |
| $n$ | 0.72 (±0.01) | 0.65 (±0.01) | 0.73 (±0.01) | 0.70 (±0.01) | 0.73 (±0.01) | 0.70 (±0.01) |
| $C_1^*$/mF cm$^{-2}$ [2] | 7.5 | 12.8 | 6.7 | 11.4 | 7.1 | 12.1 |
| $\chi^2$ | 0.013 | 0.028 | 0.011 | 0.014 | 0.009 | 0.022 |

(1) Steady-state cell temperature attained before data acquisition by self-heating (ambient conditions: 22 °C, 30 % RH).

(2) Effective capacitance for a normal to the surface time constant dispersion [47,48]: $C_1^* = Y_1^{1/n} \cdot R_1^{(1/n-1)}$

**Table II.** Results of the impedance spectroscopy analysis (Fig. 5).

More insights into liquid water behavior inside the passive PEMFC can be gained with neutron radiography.

*3.2 Neutron radiography*

Neutron radiography was carried out on the passive PEMFC in vertical and horizontal orientations. For the vertical cell, images were acquired with frontal and lateral incidence. The $H_2$ inlet port position (up or down) was also tested in the vertical cell to determine any possible influence of the $H_2$ flow inside the anodic chamber on water distribution. Images of the horizontal cell were obtained under lateral neutron incidence, with cathode-up and cathode-down orientations.

**Fig. 3** shows neutron radiographs of a cell in vertical position and under frontal incidence, with $H_2$ feeding through the down-port (Fig. 3a) and up-port (Fig.3b). Water appears as a spotted white contrast inside the active area of the cell. The amounts of water generated at the acquisition of the images are indicated in mg cm$^{-2}$. The evolution of cell parameters (current, voltage, power, energy, water generation, and temperature) during the acquisition of the images is given in Supplementary Information, S4, Figs. S6 and S7.

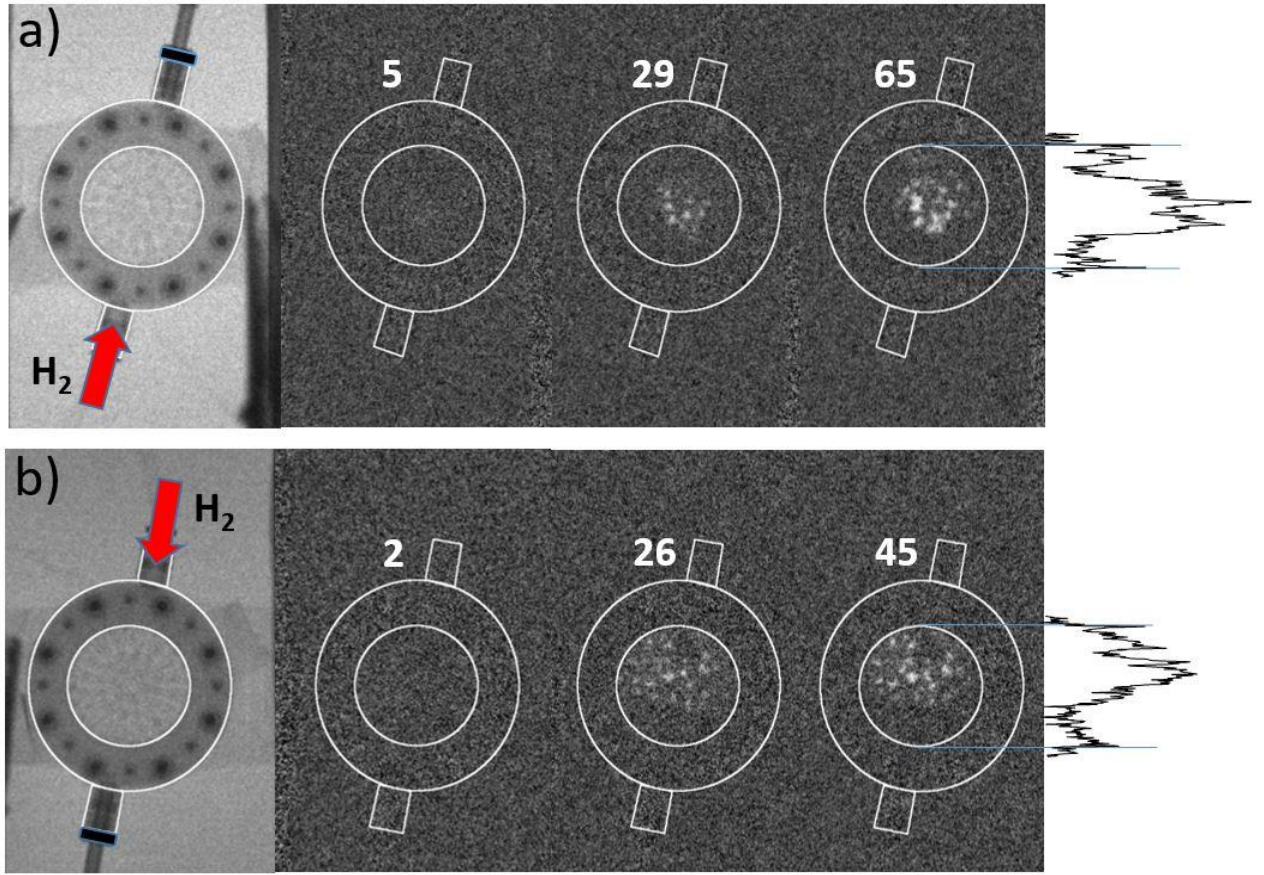

**Fig. 3**. Thermal neutron radiography under frontal incidence of the passive PEMFC in vertical position, with $H_2$ inlet from (a) down-port, and (b) up-port. The left image is the dry reference image ($IM_{dry}$, Eq. 2). Numbers indicate the amount of water produced by the cell in mg cm$^{-2}$. Cell active area, frame, and $H_2$ ports are drawn for visual guidance. Right: pixel values profiles taken over the center diameter of the active area.

The neutron images of Fig. 3 show liquid water accumulations preferentially centered in the active area of the cell with a slightly higher concentration in the upper half. The plotted distribution of water resembles the columns array of the cathodic plate, which appear to favor the condensation due to more hydrophilic and colder conditions than the electrode surface. In a similar way, preferential water accumulation is observed under the lands of flow-field channels in conventional cells [38]. The deviation of water condensation to the upper half of the active area can be explained by the combined effects of gravity and natural convection. The water droplets located in the lower half will leave the cell faster, also entrained by larger drops falling down from the upper half of the cell. Natural convection contributes to this asymmetry, since ambient air rising along the hot fuel cell will first "dry" the lower part of the cell thus reaching the upper half with a high vapor-content. The effect of natural forces on water distribution and passive cell performance is further discussed in the following section. The position of the inlet port, on the other hand, does not appear to include a significant difference in liquid water distribution.

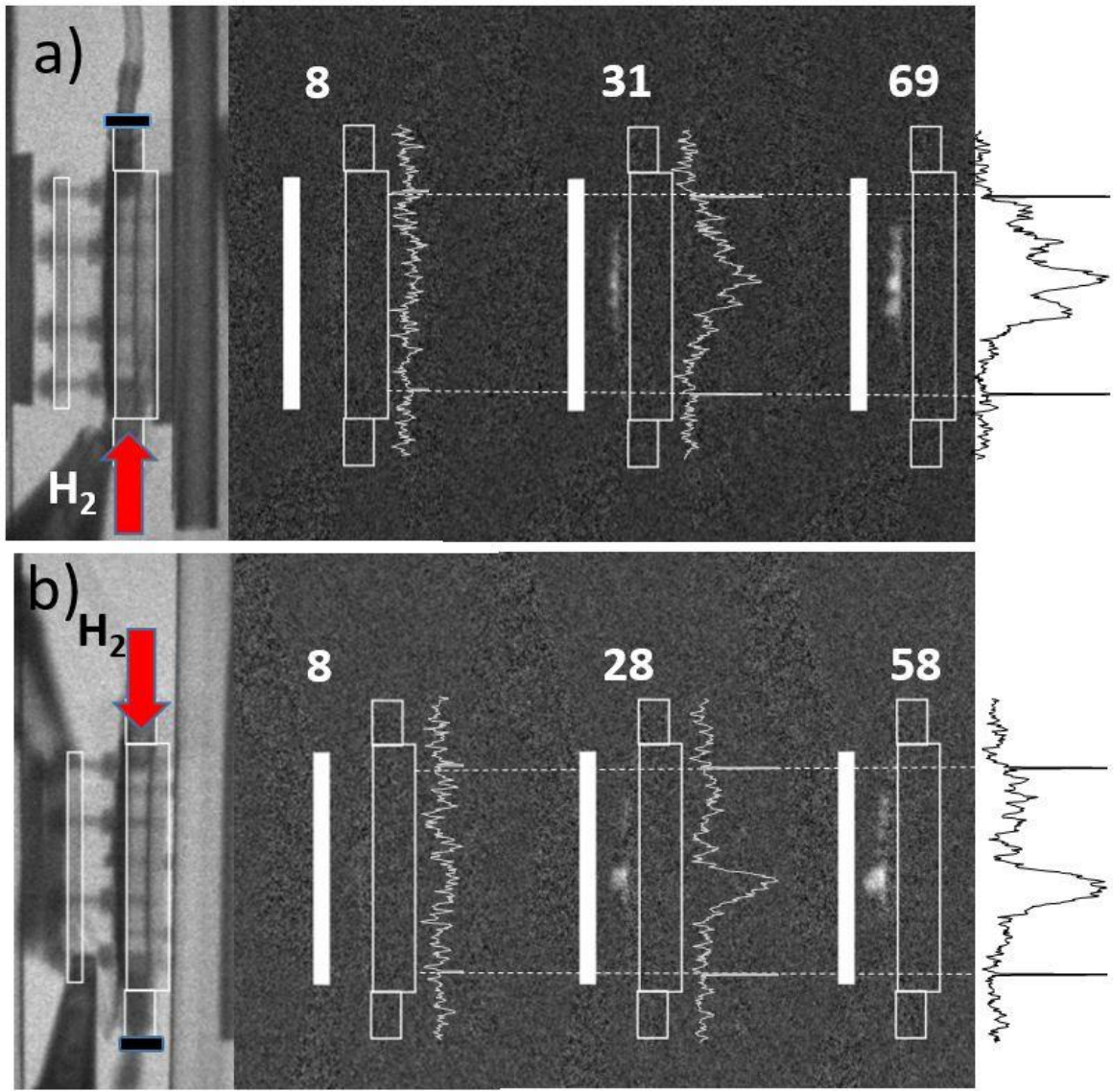

**Fig. 4**. Thermal neutron radiography images, and gray scale profiles, under lateral incidence in a passive PEMFC operating in vertical position, with (a) $H_2$ inlet from down-port, and (b) $H_2$ inlet from up-port. Numbers indicate the amount of water produced by the cell in mg cm$^{-2}$. Cathode plate (white bar), anode plate (empty bar), and $H_2$ ports are drawn for visual guidance. The left image is the dry reference image ($IM_{dry}$, Eq. 2).

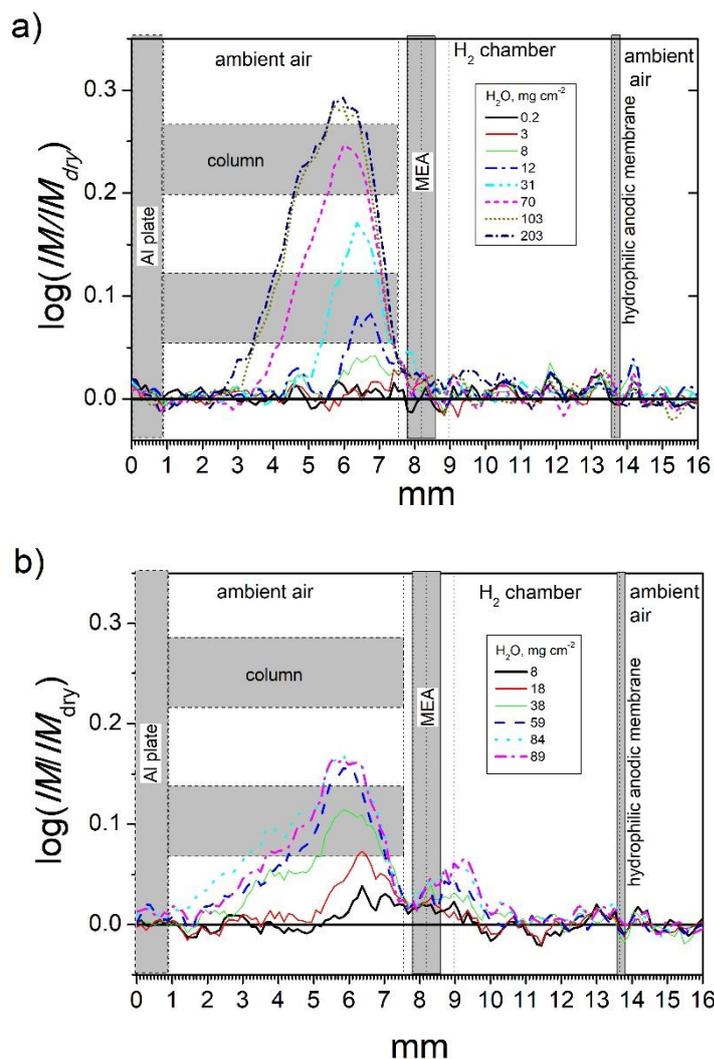

**Fig. 5**. Gray intensity vs. cross sectional distance in a passive PEMFC operating in vertical position, with $H_2$ inlet from (a) down-port, and (b) up-port, taken from neutron images in Fig. 4. Gray areas indicate the main components of the cell.

Images of the vertical cell under lateral neutrons exposure are shown in **Fig. 4**, with the two different $H_2$ port locations (See Supplementary Information S4, Figs. S8 and S9, for cell parameters evolution during the acquisition). As in the frontal incidence images (c.f. Fig. 3), the lateral images and gray scale profiles reflect preferential water accumulation in the central part of the active area, with larger accumulation in the upper half. Water concentration profiles as a function of cross sectional distance in the cell are shown in **Fig. 5**, by averaging gray levels over vertical slices covering the active area section from Fig. 4 images.

The profiles show the preferential water accumulation around the columns of the cathodic plate, reflecting the passive water removal from the cathode surface by the more hydrophilic and colder cathodic plate. Some water accumulation is observed in the anode grid contact (grids are plotted as vertical dotted line in Fig. 5), especially in Fig. 5b, with $H_2$ feeding from the upper port. The MEA region, right in the center of the profile, shows an apparent lower water concentration which must reflect the decrease in absolute volume concentration inside saturated porous media under operation (c.f. Table I), although a change in the sensitivity of neutrons cannot be discarded here. The hydrophilic anodic membrane, which is placed behind the anode grid to help passive water removal, shows minor water accumulation which may be due to effective permeation. A steady-state profile is attained after about 70 mg cm$^{-2}$ water generation which corresponds with the stabilization of cell voltage and temperature at current density above 0.2 A cm$^{-2}$ and stable power production (Supplementary Information, S4, Figs. S8 and S9).

A different situation occurs when the cell is in horizontal position (**Fig. 6**). Two cases have been analyzed, cathode-up (Fig. 6a), and cathode-down (Fig. 6b). (See Supplementary Information, S4, Figs. S10 and S11, for cells parameter evolution during images acquisition).

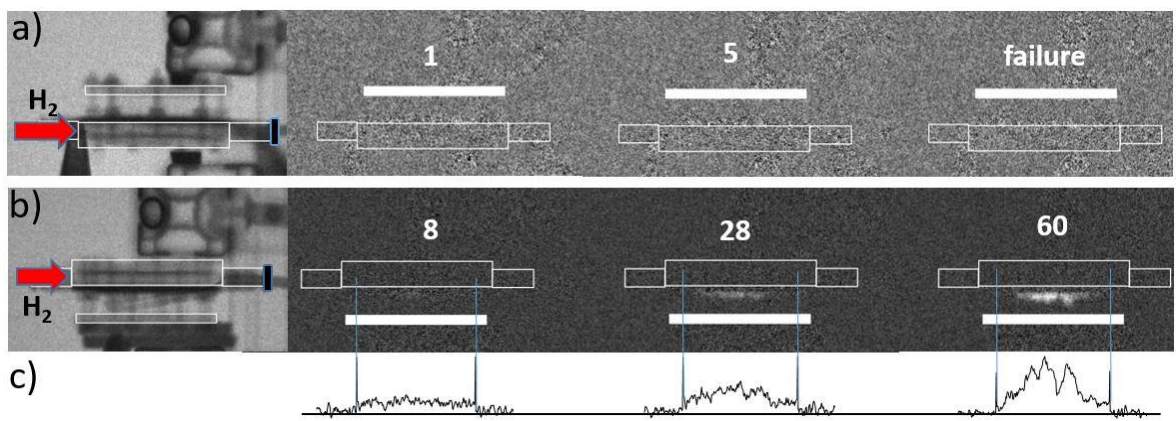

**Fig. 6**. Thermal neutron radiography images under lateral incidence in a passive PEMFC operating in horizontal position, with (a) cathode-up, and (b) cathode-down. Numbers indicate the amount of water produced by the cell in mg cm$^{-2}$. Cathode plate (white bar), anode plate (empty bar), and H$_2$ ports are drawn for visual guide. The left image is the dry reference image ($IM_{dry}$, Eq. 2). c) Plots of the gray intensity over the active area in b.

Cathode-up gives rise to a drop in voltage and cell failure shortly after production of 5 mg cm$^{-2}$ of water. As a result, no water accumulation can be observed in the neutron images with cathode-up. The cell is returned to full operative state after a short anode purging. With cathode-down, however, the cell operates without failure; plots of the gray intensity over the active area (Fig. 6c) show initially rather uniform water generation profile (8 mg cm$^{-2}$ water production), followed by preferential accumulation of large drops over central region of the cathodic plate after 60 mg cm$^{-2}$ water production, when a steady-state cell voltage and power generation are attained. Random oscillations in cell voltage observed in Fig. S11, with the amplitude increasing with current density, reflect a large effect of water drops emerging and moving over the surfaces of the cathode on cell response, most probably by changing oxygen accessibility. They show that the response of the horizontal cell is more sensitive to water dynamics over the cathode surface than in vertical position (cf. Figs. S6-S11). Cross sectional water profiles extracted from Fig. 6 are plotted in **Fig. 7**.

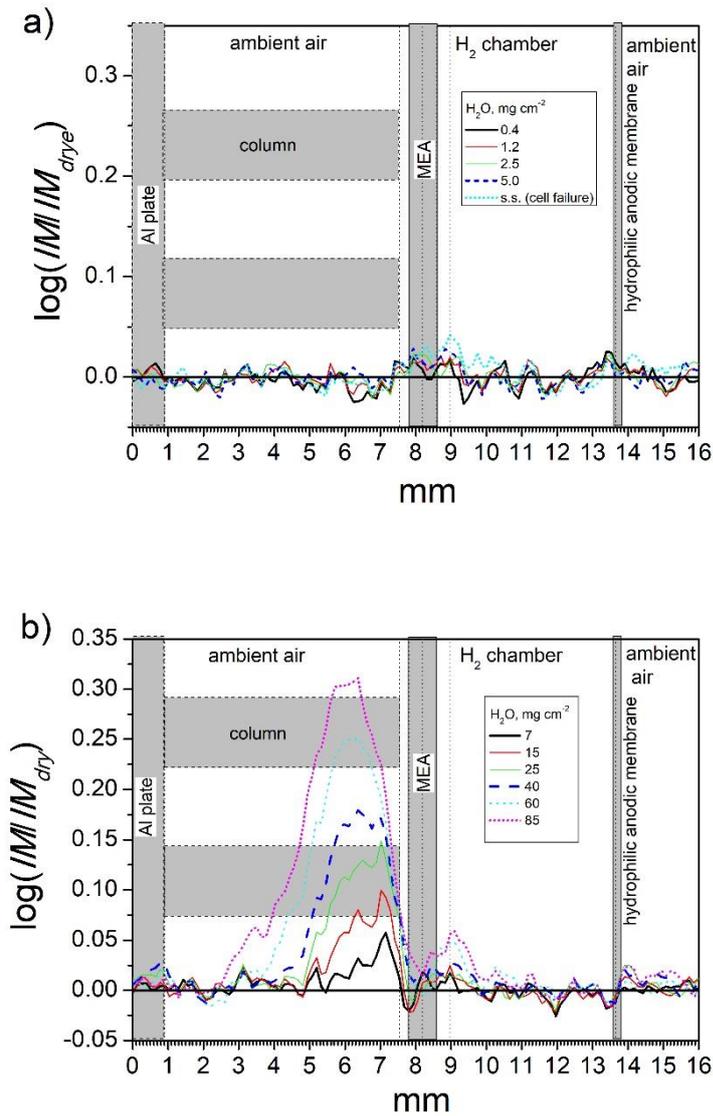

**Fig. 7**. Gray intensity vs. cross sectional distance in a passive PEMFC operating in horizontal position, with (a) cathode up, and (b) cathode down. Gray areas indicate the main components of the cell.

The profile in Fig. 7 show little water generation and accumulation with cathode-up due to early failure of this cell after the production of 5 mg cm$^{-2}$. With cathode-down, no failure is observed and a water profile evolves in the cathode and anode apparently in a similar way as for the vertical cell (c.f. Fig. 5) in spite their significant differences in polarization curves (Fig. 2) and impedance spectroscopy (Fig. S5). A closer comparison is carried out in the next section, that shows essential differences among profiles that explain the different cell response.

## 4. Discussion

*4.1 Power generation and passive PEMFC orientation*

Polarization curves and impedance spectroscopy show an important influence of cell orientation on the performance of the passive PEMFC. There are some studies of the effect of cell orientation on air breathing PEMFC performance, showing different results. Jang et al. [49] report an increase in cell performance in vertical position, using a cell with vertical slits in the cathodic plate. Obeisun et al. [50] study an air breathing cell with long horizontal openings in the cathodic plate, and observed that cell orientation has little effect, only apparent after a prolonged operation time. The different results must be attributed to different designs used for the cathodic plate in these two works. A plate that allows droplets movement over the cathode surface, as in the design by Jang et al. [49], will make the cell more sensitive to orientation and with a favored vertical position, whereas the design of Obeisun et al. [50] is more prone to accumulate more static water drops on the cathode surface that are preferentially removed by evaporation. Together with the shape of the openings, most determining for air-breathing performance are the thickness of the perforated cathodic plate, that limits water drops movement and cell performance [51], and the gas diffusion properties of the electrode [52]. Cell orientation has also been analyzed with theoretical models of air-breathing cells, showing their thermal dissipation and response favored in vertical position by natural convective heat flow [53,54].

For the design in Fig. 1, the cathodic columnar plate is characterized by null thickness in contact with the cathode surface which allows for a high mobility of water drops. This property leads to lower mass transport resistance and larger peak power density when the cell is in vertical position (Fig. 2). The changes in water contents and mass transport resistance reflect a high orientation dependence of the two principal natural forces, namely the gravitational and natural convection. Gravitational rolling of water over the electrode surface in vertical position is important when single drops are large enough to overcome adhesion by surface tension effects, which pin them to the cathode surface, current collector grid, and the columnar cathodic plate. Such water droplet removal is facilitated by hydrophobic elements that lower the adhesive forces. The natural convection caused by the "hot" fuel cell and oxygen/vapor concentration gradients [55] might even have more significant impact than gravitational dragging on liquid water removal. This phenomenon has been modelled by Ismail et al. [41], and experimentally observed by Fabian et al. [56] and Coz et al. [23]. For vertically aligned cells, a cold fluid particle rising in front of the hot GDE will be heated more and more as it rises, draining air from the surroundings and forming a strong upwards directed jet that enhances water evaporation from the cathode surface. The horizontally aligned fuel cell, however, resembles the classical Rayleigh-Bénard instability, where an air layer is heated from below, but natural convection will hardly reach the GDEs. As a result, the horizontal cell does not benefit from improved dragging of water and natural convection over the cathode surface.

The performance of a passive air-breathing PEMFC, with a high water mobility over the cathode surface, is more dependent on orientation than a conventional cell, where formation of small droplets on the surface of the gas diffusion layer and their dragging in the channels by the gas stream minimizes the effects of gravity and natural convection [57].

*4.2 Water distribution in the passive PEMFC and cell performance*

Water distribution profiles obtained from neutron radiography explain the differences in cell performance from vertical to horizontal positions. In spite of the similarity in water profiles of the vertical and horizontal cathode-down positions observed from Figs. 5b and 7b, a closer view, almost in the limit of the resolution of the technique, allows to see determinant differences for cell performance.

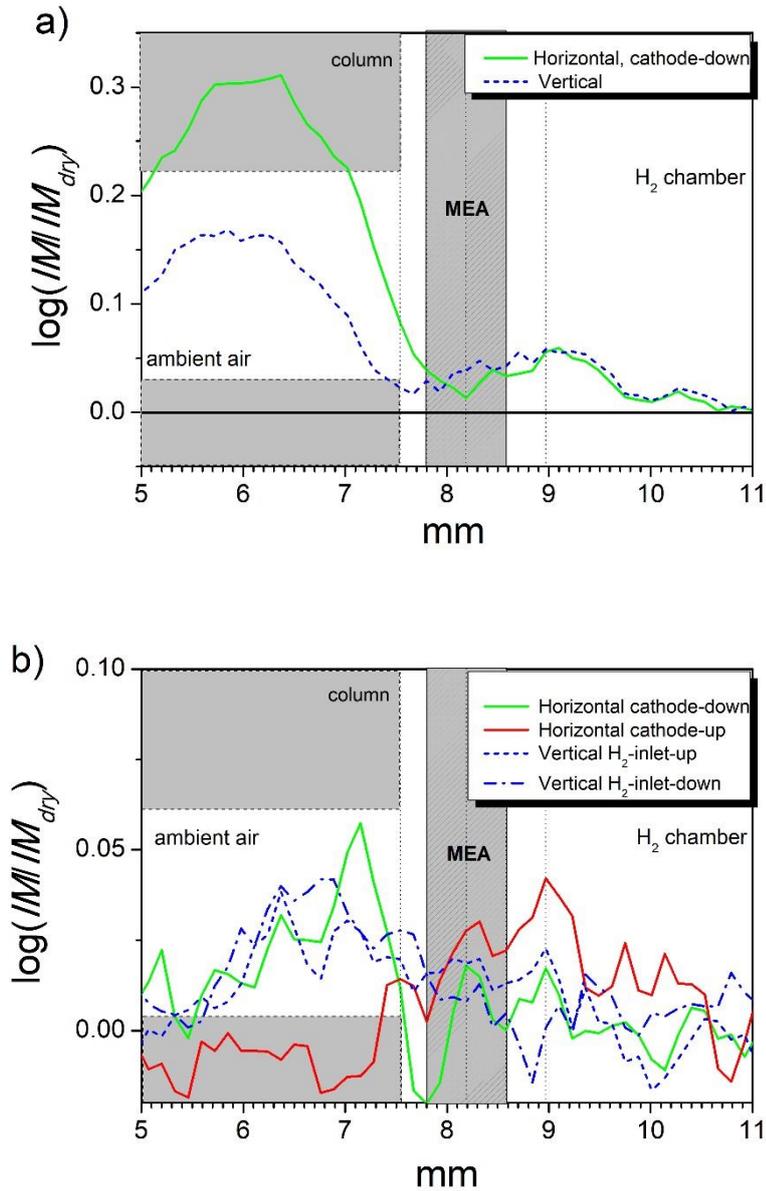

**Fig. 8**. Comparing steady-state water profiles, after production of 85 mg cm$^{-2}$ H$_2$O (a), and after production of 5 mg cm$^{-2}$ H$_2$O (b), for the indicated cell orientations.

**Fig. 8a** shows the steady-state water profiles in the MEA region for a vertical and a horizontal cathode down cell. The steep rise in water profile from the cathodic side of the MEA, towards the cathode surface and columnar plate, reflects larger amount of water inside the cathode of the horizontal cell, in accordance with the analysis of polarization curves (Table I). In the anode side, however, the water profiles are very similar for both orientations.

A change in water distribution from cathode to anode, and dependent with cell orientation, reflects different nature of forces acting for water transport. In the cathode, water transport is dominated by gravity and natural convection over the surface, which are very dependent on cell orientation. In the anode, water is driven by diffusion through the membrane and capillarity in the porous electrode, rather insensitive to cell orientation. In fluid-dynamics terms, the passive air-breathing PEMFC is characterized by different Bond number in cathode and in anode (the Bond number, or Eötvös number, describing the relative importance of gravitational forces and surface tension). Whereas the anode accomplishes Bo << 1, as in a conventional PEMFC electrode [57], for the air-breathing cathode the situation is Bo > 1.

Neutron images allow to explain the conditions for cell failure observed with horizontal cell position and cathode up, see also Fig. S10, after production of a small volume of water (35·$10^{-3}$ $cm^3$) which amounts approximately to 20 % of the open pores volume of the cathode. A deleterious effect by such small amount of water can only be explained by the flooding of a small, but critical, part of the MEA like the anodic catalyst layer. This situation is inferred from a comparison of the neutron water profiles for the different cell orientations in the MEA region, and after production of 5 mg $cm^{-2}$ of water (**Fig. 8b**), showing that the horizontal cell with cathode-up is characterized by a larger water concentration in the anode side than in the cathode side, as a difference from the other cells. The cathode on top increases the flow of water to the anode, causing the flooding of the catalyst layer, which is critical for cell operation, most probably due to the low solubility of hydrogen in water, and leads to immediate failure.

## 5. Conclusions

The main conclusions obtained from the experimentation and analysis shown in this work are:

- Thermal neutron radiography at the AKR-2 (2 W) is able to resolve build-up of liquid water in the cathode and the anode of a passive air-breathing PEMFC.

- Liquid water accumulation is predominant of the cathode surface, but with small differences depending on cell orientation that determine cell response.

- The response of the cell shows 17 % larger peak power density in vertical with respect to horizontal position.

- Polarization curves analysis and impedance spectroscopy show that the horizontal cell presents larger mass transport losses due to higher water contents in the cathodic gas diffusion layer. Neutron imaging confirms more important water accumulation in the cathodic electrode of the horizontal cell.

- Water saturation and performance changes with cell orientation are attributed to the different action of two main natural forces: gravity and natural convection. The columnar cathodic plate used allows for high mobility of water drops over the cathode surface and favors cell response in vertical position.

- Laying the cell horizontal with the cathode up produces eventual failure. In this position, back-diffusion and gravity operate in the same direction leading to flooding of the anodic catalyst layer.

**Acknowledgements**

This work was supported by the Ministerio de Economía y Competitividad of Spain, and Fondo Europeo de Desarrollo Regional (FEDER), Project E-LIG-E, ENE2015-70417-P (MINECO/FEDER).

**Appendix 1**

A polarization curve for a passive air-breathing PEMFC which accounts for oxygen diffusivities in the cathodic electrode is described in Ref. 35. The latter is based on a one dimensional model for a flooded agglomerate porous catalyst layer, which considers both charge transfer kinetics and oxygen diffusion losses within pores. The model assumes macroscopically homogeneous electrode layers, the gas diffusion layer and the catalyst layer, with uniform overpotential distribution over the electrode thickness; the cathode kinetics is governed by first order oxygen reduction with Tafel behavior. The polarization curve obeys the general expression in Eq. A1 with three main potential losses, electrochemical ($\eta_{ele}$), ohmic ($\eta_{ohm}$), and mass transport ($\eta_{mt}$):

$$V = E - \eta_{ele} - \eta_{ohm} - \eta_{mt} \tag{A1}$$

Expressions for each overpotential term are determined as a function of current density. The electrochemical overpotential ($\eta_{ele}$) is due to the electrochemical reaction in the cathodic catalyst layer (no anodic polarization is considered). To calculate its relation with current density and oxygen diffusivity, the following expression is used:

$$D_{CL} \frac{d^2 c(x)}{dx^2} = \frac{i(x)}{nF} \tag{A2}$$

Here, $c(x)$ is the local oxygen concentration as a function of depth ($x$, see Eq. A4) in the catalyst layer, $i(x)$ the local current density, $D_{CL}$ is the effective diffusion coefficient of oxygen in the catalyst layer, $n(= 4)$ the number of exchanged electrons per oxygen molecule, and $F(= 96485$ F mol$^{-1}$) the Faraday constant. The local current density ($i(x)$) depends on local oxygen concentration, electrochemical overpotential, and other kinetics parameters according to:

$$i(x) = \frac{c(x)}{c^0} A_i \, i_0 \, exp\left(\eta_{ele} \frac{\alpha \, nF}{RT}\right), \tag{A3}$$

where $A_i$ (cm$^2$ cm$^{-3}$) is the internal electrochemical area of the catalyst layer, $i_0$ (A cm$^{-2}$) the exchange current density, and $\alpha$ the transfer coefficient. Notice that positive signs are used for the overpotentials (as in Eq. A1) and for the reduction current; therefore the exponential term in Eq. A3 is also positive.

Eq. A2 is solved with boundary conditions at the membrane/CL (Eq. A4a) and CL/GDL (Eq. A4b) interfaces:

$x = L_{CL}$     $dc/dx = 0$     (A4a)

$x = 0$     $c = c^0$ ,     (A4b)

where $L_{CL}$ is the thickness of the catalyst layer, and $c^0$ is the oxygen concentration at the CL/GDL interface. Substitution of Eq. A3 into A2 and integration gives for the concentration profile:

$$c(x) = c^0 \left(\frac{exp(2L_{CL}\sqrt{a})+exp(2x\sqrt{a})}{1+exp(2L_{CL}\sqrt{a})}\right) exp(-x\sqrt{a}),$$     (A5)

where:

$$a = \frac{A_i\, i_0}{nFD_{CL}c^0}\ exp\left(\eta_{ele}\frac{\alpha\, n\, F}{RT}\right)$$     (A6)

The parameter $1/a^{1/2}$ is the penetration depth of the electrochemical reaction within the catalyst layer. The current profile ($i(x)$) and total current density ($i_t$) are given by:

$$i(x) = nF\, D_{CL} a\, c(x)$$     (A7)

$$i_t = nF\, D_{CL} a\, \int_0^{L_{CL}} c(x)d(x)$$     (A8)

Integrating Eq. A8 gives for the total current:

$$i_t = nFD_{CL}c^0 a^{1/2} tanh(a^{1/2}L_{CL})$$     (A9)

The concentration at the CL/GDL interface, $c^0$, is obtained from the expression for a linear diffusion process in terms of gas diffusion layer parameters:

$$i_t = nF\, D_{GDL}\frac{c^*-c^0}{L_{GDL}} ,$$     (A10)

where $c^*$ is the oxygen concentration at the GDL/air, $L_{GDL}$ the gas diffusion layer thickness, and $D_{GDL}$ the effective oxygen diffusion coefficient in the GDL. For $c^*$, the dissolved concentration in water will be used depending on oxygen partial pressure ($p_{O2}$) through Henry isotherm:

$$c^* = k_H \, p_{O2} \tag{A11}$$

Here $k_H$ (=1.2·10$^{-6}$ mol cm$^{-3}$ atm$^{-1}$) is the Henry isotherm constant of oxygen in water. Taking $c^0$ from Eq. A10 and substituting into Eqs. A6 and A9:

$$a = \frac{A_i \, i_0}{nFD_{CL}\left(c^* - i_t \frac{L_{GDL}}{nF\,D_{GDL}}\right)} \, exp\left(\eta_{ele} \frac{\alpha \, nF}{RT}\right) \tag{A12}$$

$$\boldsymbol{i_t} = nF \, D_{CL} \frac{c^* \sqrt{a}}{coth(L_{CL}\sqrt{a}) + \frac{D_{CL}}{D_{GDL}} L_{GDL} \sqrt{a}} \tag{A13}$$

Eqs. A12 and A13 form a system of equations with two unknowns ($a$, $i_t$), and with $\eta_{ele}$ as parameter, which can be solved by iterative procedure, using $c^0 = c^*$ in Eq. A6 for the seminal $a$ value.

The voltage loss due to mass transport within the GDL is expressed as a function of the current density and the diffusion limiting current ($i_L$):

$$\eta_{mt} = \frac{RT}{\alpha \, nF} log\left(\frac{i_L}{i_L - \boldsymbol{i_t}}\right) \tag{A14}$$

Finally, the ohmic overpotential is calculated according to:

$$\eta_{ohm} = \boldsymbol{i_t} \, R_{ohm} \tag{A15}$$

where the ohmic resistance ($R_{ohm}$) is due to electric contacts, specially between the grid collector and the cathode ($R_{cont}$), and the ionic resistances of the membrane and catalyst layers ($R_{mem}$, $R_{CL}$, respect.): $R_{ohm} = R_{CL} + R_{mem} + R_{cont}$

The thermodynamic potential in Eq. A1 ($E$) can be calculated as a function of temperature ($T$) and partial pressures of hydrogen and oxygen ($p_{H2}$, $p_{O2}$) from:

$$E = 1.229 - 0.85 \, 10^{-3}(T - 298.15) + RT log(p_{H2}^2 \, p_{O2}) \tag{A16}$$

Finally, the polarization curve in Eq. A1 results:

$$V = E - \eta_{ele} - \boldsymbol{i_t} \, R_i - \frac{RT}{\alpha \, nF} log\left(\frac{i_L}{i_L - \boldsymbol{i_t}}\right) \tag{A17}$$

The fitting to an experimental curve is carried out by first calculating the total current density, $\boldsymbol{i_t}$, for each $\eta_{ele}$ by using the system of Eqs. A12 and A13, and then the other overpotentials and the cell voltage from $\boldsymbol{i_t}$ to obtain the polarization curve from Eq. A17.

The calculated diffusivities of the gas diffusion layer and catalyst layers, $D_{GDL}$ and $D_{CL}$, for the three cell orientations, together with Tafel slope, $b$, and internal resistance, $R_i$, are given in Table A1.

| Cell orientation | $10^4 \cdot D_{CL}$ cm² s⁻¹ | $10^3 \cdot D_{GDL}$ cm² s⁻¹ | $b(=2.303RT/\alpha nF)$ V dec⁻¹ | $R_i$ Ohm cm² | $10^4 \cdot X_i^2$ |
|---|---|---|---|---|---|
| Vertical | 8.0 | 145 | | | 5.5 |
| Horizontal, cathode down | 6.7 | 109 | 0.097 | 0.45 | 2.5 |
| Horizontal, cathode up | 6.5 | 111 | | | 2.6 |

Table A1.

Parameters $b$ and $R_i$ were fitted for the vertical cell and their values fixed for the other two cell positions, since they are considered not to change significantly with cell orientation. Other common parameters used are:

$A_i$, electrocatalyst area of the cathodic catalyts layer, 60000 cm² cm⁻³ (measured by the hydrogen desorption technique)

$c^*(=k_{Henry} \, p_{O2})$, oxygen dissolved concentration in water, $0.25 \cdot 10^{-6}$ mol cm⁻³

$E^0$, thermodynamic potential, 1.22 V (Eq. A.16)

$i_0$, exchange current density in cathode, $8.5 \cdot 10^{-9}$ A cm$^{-2}_{Pt}$

$L_{CL}$, cathodic catalyst layer thickness, 15 μm (measured SEM)

$L_{GDL}$, cathodic gas diffusion layer thickness, 400 μm (measured SEM)

$p_{H2}$, hydrogen pressure in anode, 1.5bar$_g$

$p_{O2}$, oxygen partial pressure in cathode, 0.21 bar$_g$

$T$, cell temperature, 298 K